# アプリケーション機能ブロックのGPU，FPGA自動オフロード手法の評価

山登庸次†

† NTT ネットワークサービスシステム研究所，東京都武蔵野市緑町 3-9-11
E-mail: †yoji.yamato.wa@hco.ntt.co.jp

**あらまし**　この4，5年で，CPUに比べた電力効率等の長所からFPGA，GPU等を利用したシステムが増えている．しかし，FPGA，GPU等のシステムでの利用には，HDLやCUDA等のハードウェアを意識した技術仕様の理解が必要であり，ハードルは高い．これらの背景から，私は，プログラマーがCPU向けに開発したソースコードを，適用される環境に応じて，自動で変換し，リソース量等を設定して，高い性能で運用可能とする環境適応ソフトウェアのコンセプトを提案している．そのコンセプトの要素として，CPU向けアプリケーションソースコードのループ文を，FPGA，GPUに自動オフロードする方式を提案評価している．本稿では，GPU，FPGAへの自動オフロードでより高速化を実現するため，アプリケーションの中で個々のループ文でなくより大きな単位である機能ブロックをオフロードする手法について，提案，評価を行う．提案手法を既存アプリケーションで評価する．

# Evaluation of Automatic GPU and FPGA Offloading for Function Blocks of Applications

Yoji YAMATO†

† Network Service Systems Laboratories, NTT Corporation, 3-9-11, Midori-cho, Musashino-shi, Tokyo
E-mail: †yoji.yamato.wa@hco.ntt.co.jp

**Abstract**　In the recent years, systems using FPGAs, GPUs have increased due to their advantages such as power efficiency compared to CPUs. However, use in systems such as FPGAs and GPUs requires understanding hardware-specific technical specifications such as HDL and CUDA, which is a high hurdle. Based on this background, I previously proposed environment adaptive software that enables automatic conversion, configuration, and high--performance operation of once written code according to the hardware to be placed. As an element of the concept, I proposed a method to automatically offload loop statements of application source code for CPU to FPGA and GPU. In this paper, I propose and evaluate a method for offloading a function block, which is a larger unit, instead of individual loop statements in an application, to achieve higher speed by automatic offloading to GPU and FPGA. I implement the proposed method and evaluate with existing applications offloading to GPU.

## 1. は じ め に

近年，CPUの半導体集積度が1.5年で2倍になるというムーアの法則が減速するのではないかと言われている．そのような状況から，メニーコアのCPUだけでなく，FPGA（Field Programmable Gate Array）やGPU（Graphics Processing Unit）等のハードウェアの活用が増えている．例えば，Microsoft社はFPGAを使ってBingの検索効率を高めるといった取り組みをしており[1]，Amazon社は，FPGA, GPU等をクラウドのインスタンス（例えば，[2]-[14]）として提供している[15].

しかし，CPU以外のハードウェアをシステムで適切に活用するためには，ハードウェアを意識した設定やプログラム作成が必要であり，OpenCL（Open Computing Language）[16]，CUDA（Compute Unified Device Architecture）[17]といった

知識が必要になってくるため，大半のプログラマーにとっては，スキルの壁が高い．

一方，IoT（Internet of Things）技術（例えば，[18]- [22]）は普及してきており，ネットワークにつながるデバイスも，既に数百億と増えており，2030 年には兆台がつながると予測されている．IoT を用いた応用は，医療，流通，製造，農業，エンタメ等に広がっており，サービス合成技術等 [23]- [30] を活用して，製品が届くまでの過程を可視化するなどの応用がされている．

IoT を用いたシステムで，IoT デバイスを詳細まで制御するためには，組み込みソフトウェア等のスキルが必要になることがある．Raspberry Pi 等の小型端末をゲートウェイ（GW）に，多数のセンサデバイスを集約管理することも頻繁にされるが，小型端末の計算リソースは限定されるため，利用環境に応じて管理の設計が必要となる．

背景を整理すると，CPU 以外の GPU や FPGA 等のハードウェア，多数の IoT デバイスを活用するシステムは今後ますます増えていくと予想されるが，それらを最大限活用するには，壁が高い．そこで，そのような壁を取り払い，CPU 以外のハードウェアや多数の IoT デバイスを十分利用できるようにするため，プログラマーが処理ロジックを記述したソフトウェアを，配置先の環境（FPGA，GPU や IoT GW 等）にあわせて，適応的に変換，設定し，環境に適合した動作をさせるような，プラットフォームが求められている．

Java [31] は 1995 年に登場し，一度記述したコードを，別メーカーの CPU を備える機器でも動作可能にし，環境適応に関するパラダイムシフトをソフト開発現場に起こした．しかし，移行先での性能については，適切であるとは限らなかった．そこで，私は，一度記述したコードを，配置先の環境に存在する GPU や FPGA，IoT GW 等を利用できるように，変換，リソース設定等を自動で行い，アプリケーションを高性能に動作させることを目的とした，環境適応ソフトウェアを提案した．合わせて，環境適応の要素として，アプリケーションソースコードのループ文を，FPGA，GPU に自動オフロードする方式を提案評価している [32] [33]．本稿では，GPU，FPGA への自動オフロードでより高速化を実現するため，アプリケーションの中で個々のループ文でなくより大きな単位である機能ブロックをオフロードする手法について，提案，評価を行う．提案手法を既存アプリケーションで有効性を評価する．

## 2. 既存技術

環境適応ソフトウェアとしては，Java がある．Java は，仮想実行環境である Java Virtual Machine により，一度記述した Java コードを再度のコンパイル不要で，異なるメーカー，異なる OS の CPU マシンで動作させている（Write Once. Run Anywhere）．しかしながら，移行先で，どの程度性能が出るかはわからず，移行先でのデバッグや性能に関するチューニングの稼働が大きい課題があった（Write Once, Debug Everywhere）．

GPU の並列計算パワーを画像処理でないものにも使う GPGPU（General Purpose GPU）（例えば [34]）を行うための環境として CUDA が普及している．CUDA は GPGPU 向けの NVIDIA 社の環境だが，FPGA，メニーコア CPU，GPU 等のヘテロなハードウェアを同じように扱うための仕様として OpenCL が出ており，その開発環境 [35] [36] も出てきている．CUDA，OpenCL は，C 言語の拡張を行いプログラムを行う形だが，プログラムの難度は高い（FPGA 等のカーネルと CPU のホストとの間のメモリデータのコピーや解放の記述を明示的に行う等）

CUDA や OpenCL に比べて，より簡易にヘテロなハードウェアを利用するため，指示行ベースで，並列処理等を行う箇所を指定して，指示行に従ってコンパイラが，GPU 等に向けて実行ファイルを作成する技術がある．仕様としては，OpenACC [37] や OpenMP 等，コンパイラとして PGI コンパイラ [38] や gcc 等がある．OpenACC は，Fortran/C/C++向けの仕様であるが，Java 向けには，IBM の Java JDK [39] が，Java のラムダ記述に従った GPU オフロード処理を行える．

CUDA，OpenCL，OpenACC 等の技術仕様を用いることで，FPGA や GPU へオフロードすることは可能になっている．しかしハードウェア処理自体は行えるようになっても，高速化することには課題がある．例えば，マルチコア，メニーコア CPU 向けに自動並列化機能を持つコンパイラとして，Intel コンパイラ [40] 等がある．これらは，自動並列化時に，コードの中の for 文，while 文等の中で並列処理可能な部分を抽出して，並列化している．しかし，FPGA や GPU を用いる際は，CPU と FPGA，GPU の間のメモリデータ転送のオーバヘッドのため，並列化しても性能がでないことも多い．FPGA や GPU により高速化する際には，OpenCL や CUDA の技術者がチューニングを繰り返したり，PGI コンパイラ等を用いて適切な並列処理範囲を探索し試行することがされている．

このため，OpenCL や CUDA 等の技術スキルが乏しいプログラマーが，FPGA や GPU を活用してソフトウェアを高速化することは難しいし，自動並列化技術等を使う場合も並列処理箇所の試行錯誤等の稼働が必要だった．

並列処理箇所の試行錯誤を自動化する取り組みとして，著者の以前の研究がある [32] [33]．これら研究は，GPU オフロードに適したループ文を，進化的計算手法を用いて，検証環境での性能測定を繰り返すことで，適切に抽出し，自動で高速化を行っていた．FPGA では，コンパイルに長時間かかるため，ループ文のオフロード候補を絞ってから，複数のパターンについて検証環境で性能測定する手法を提案している．しかし，特に FPGA の場合，アプリケーションに応じたハードウェア向けアルゴリズムにより高速化していることが多く，ループ文の単純なオフロードだけでは性能が不十分なことが多かった．

## 3. 機能ブロックの GPU，FPGA 自動オフロード手法の提案

### 3.1 環境適応処理のフロー

ソフトウェアの環境適応を実現するため，図 1 の処理フローを提案している．環境適応ソフトウェアは，環境適応機能を中心に，検証環境，商用環境，テストケース DB，コードパターン DB，設備リソース DB の機能群が連携することで動作する．

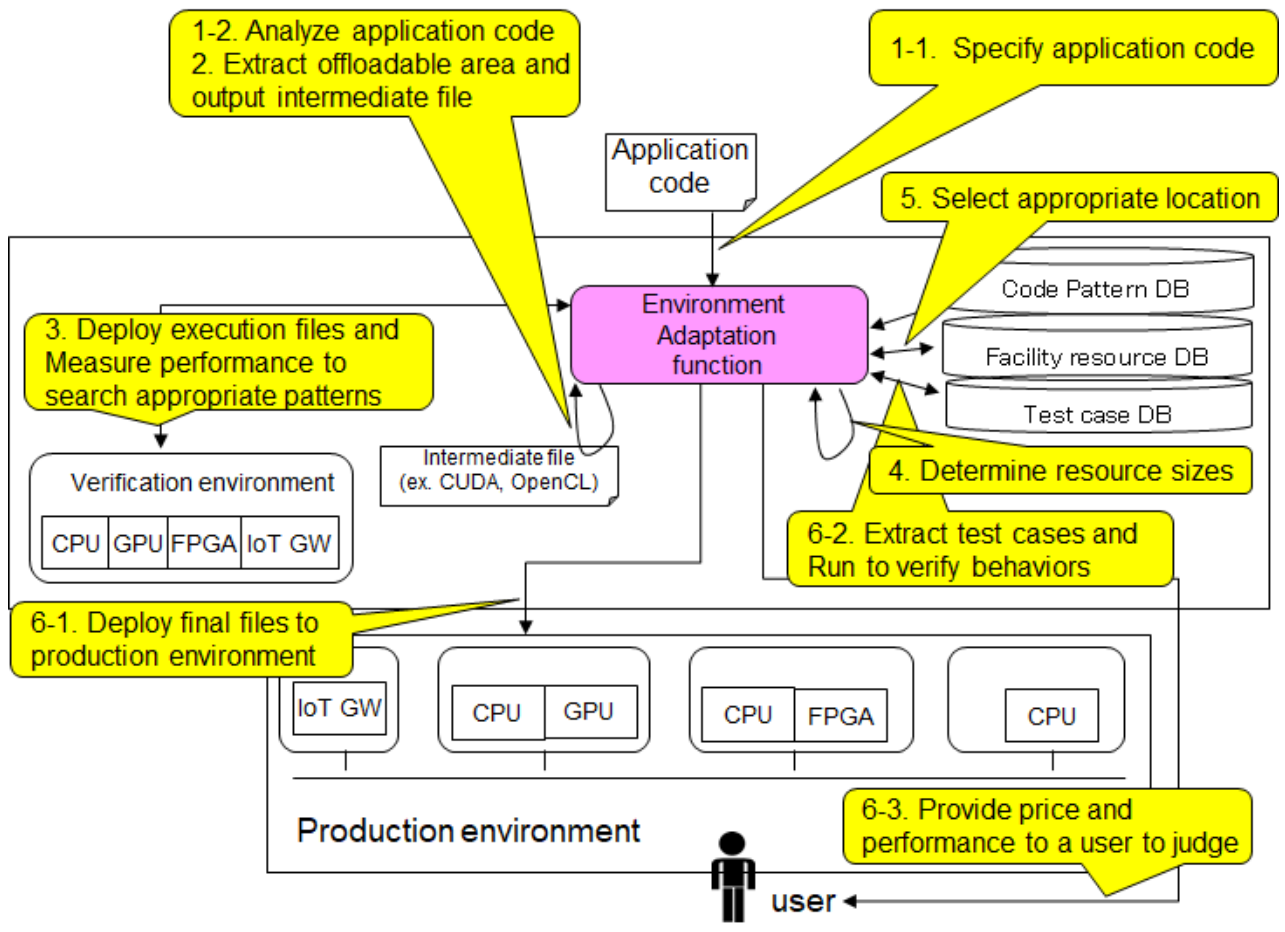

図 1　環境適応ソフトウェアのフロー

Step1 コード分析：

Step2 オフロード可能部抽出：

Step3 適切なオフロード部探索：

Step4 リソース量調整：

Step5 配置場所調整：

Step6 実行ファイル配置と動作検証：

Step7 運用中再構成：

ここで，Step 1-7 で，環境適応するために必要となる，コードの変換，リソース量の調整，配置場所の決定，Jenkins 等 [41] [42] 用いた検証，運用中の再構成を行うことができるが，実施したい処理だけ切り出すこともできる．例えば，本稿で対象とする GPU，FPGA 向けのコード変換だけ実施する場合は，Step 1-3 だけ処理すればよく，利用する機能も環境適応機能や検証環境等だけ利用する形でよい．

### 3.2　機能ブロックのオフロードの必要性

まず，著者の以前のループ文 GPU 自動オフロード手法を説明する．

GPU に適したループ文の抽出を遺伝的アルゴリズム(GA) [43] を用いて行う点を [32] は提案している．基本的な課題として，コンパイラがこのループ文は GPU で並列処理できないという制限を見つけることは可能だが，このループ文は GPU の並列処理に適しているという適合性を見つけることは難しいのが現状である．一般的にループ回数が多い等の算術強度が高いループの方が適していると言われるが，実際に GPU に出すことでどの程度の性能になるかは，実測してみないと予測は困難である．そのため，このループを GPU にオフロードするという指示を手動で行い，性能測定を試行錯誤することが行われている．

[32] はそれを踏まえ，GPU にオフロードする適切なループ文の発見を，GA で自動的に行うことを提案している．並列化を想定していない汎用プログラムから，最初に並列可能ループ文のチェックを行い，次に並列可能ループ文群に対して，GPU 実行の際を 1，CPU 実行の際を 0 と値を置いて遺伝子化し，検証環境で性能検証試行を反復し適切な領域を探索している．並列可能ループ文に絞った上で，遺伝子の部分の形で，高速化可能な並列処理パターンを保持し組み換えていくことで，取り得る膨大な並列処理パターンから，効率的に高速化可能なパターンを探索している．

また，著者の以前のループ文 FPGA 自動オフロード手法を説明する．

FPGA でも，処理時間が長くかかる特定のループ文を FPGA にオフロードして高速化することを考えた際に，どのループをオフロードすれば高速になるかの予測は難しいため，GPU 同様検証環境で試行錯誤を自動で行うことを提案している．しかし，FPGA は GPU と異なり，コンパイルに数時間以上かかるため，オフロード候補のループ文を絞ってから，実測試行を行う．発見されたループ文に対して，算術強度分析ツールを用いて算術強度が高いループ文を抽出し，更に，高算術強度のループ文に対して，展開処理等の FPGA オフロード化を行う OpenCL をプレコンパイルして，リソース効率が高いループ文を見つけ，対象を更に絞り込む．絞り込まれたループ文に対して，個々のループ文をオフロードした OpenCL やそれらループ文を組み合わせた OpenCL を生成し，FPGA 実機へコンパイルして，性能測定を行い，高速の OpenCL を解として選択する．

しかし，特に FPGA の場合，FPGA で高速化する際は，CPU 向けのアルゴリズムからハードウェア処理に適したアルゴリズムに変更し，高速化していることが多いため，ループ文の単純なオフロードだけでは，手動でアルゴリズムから変えて高速化している場合に比べ，性能が不十分なことが多かった．例えば，行列積算の場合だと，行列のすべてのデータを FPGA のローカルメモリで持つことは難しいため，データ A を行方向に読みデータ B を列方向に読んで，容量制限があるローカルメモリを上手く活用する等のアルゴリズムで高速化している例がある．GPU の場合も，フーリエ変換を GPU 向けに高速化した CUDA ライブラリの cuFFT 等は，GPU 向けのアルゴリズムで実装されている．

このようなアプリケーションに応じた処理ハードウェア向けのアルゴリズム変更は，機械に自動で抽出させるのは現状無理であるため，個々のループ文でなく，行列積算やフーリエ変換等のより大きな単位で，FPGA や GPU 等ハードウェア向けのアルゴリズム含めて実装された機能ブロックに置換することでの高速化（人の既存ノウハウの活用）を目指す．

### 3.3　機能ブロックのオフロードの処理概要と考慮点

FPGA に関しては，ハードウェア回路設計に多大な時間がかかることもあり，一度設計した機能を，IP コアという形で回路情報を利活用できるようにすることが多い．IP コアは，暗号化/復号化処理，FFT 等の算術演算，音声処理，画像処理（[44] 等）等が代表的な機能例である．IP コアはライセンス料を支払うものが多いが，一部はフリーで提供されているものもある．IP コアは開発者の既存ノウハウの塊とも言えるため，FPGA では IP コアを自動オフロードに利用することを検討する．GPU に関しては，IP コアという言い方ではないが，FFT，線形代数演算等が代表的な機能例であり，CUDA を用いて実装された cuFFT や cuBLAS 等が GPU 向けライブラリとしてフリーで提供されており，これらライブラリの活用を検討する．

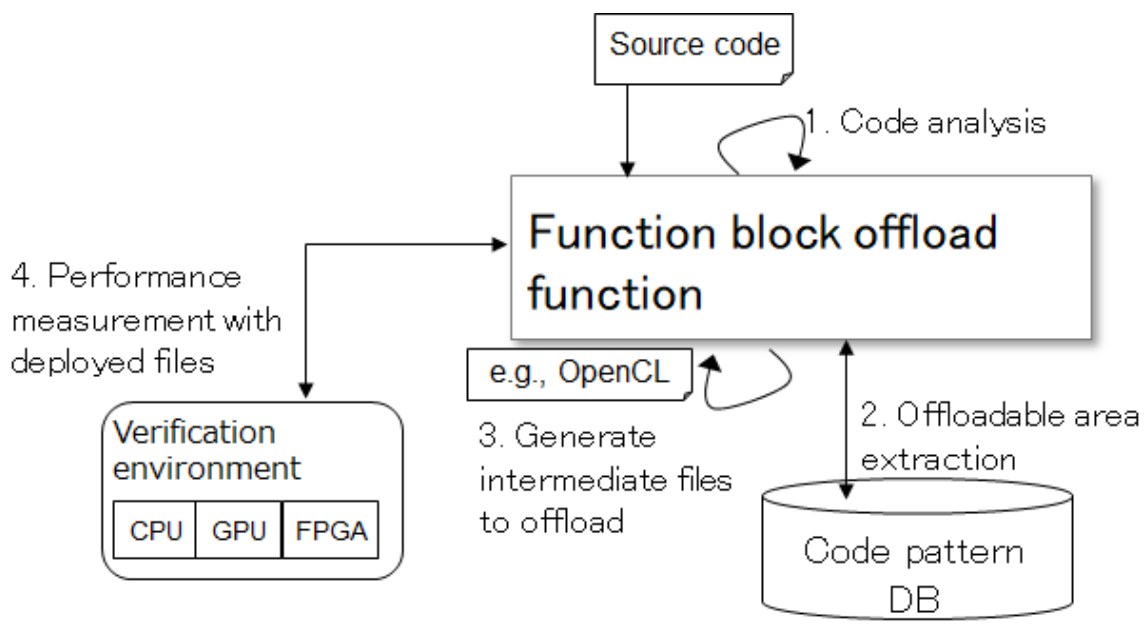

図 2　機能ブロックのオフロード処理概要

本稿では，CPU 向けに作られた既存プログラムコードの中で，FFT 処理等，GPU，FPGA にオフロードすることで高速化できるような機能ブロックが含まれる場合に，GPU 向けライブラリや FPGA 向け IP コア等に置き換えることでの高速化をする．

機能ブロックオフロードの処理概要を説明する（図 2）．Step1 にて，ソースコードの分析が行われるが，Clang 等の構文解析ツールを用いて，ループ文構造等とともに，コードに含まれるライブラリ呼び出しや，機能処理を分析する．Step1 で把握したライブラリ呼び出しや機能処理について，Step2 で，コードパターン DB と照合することで，GPU，FPGA にオフロードできる処理を発見する．Step3 で，オフロードできる処理について，GPU 向けのライブラリや FPGA 向けの IP コア等に置換して，CPU プログラムとのインタフェースを作成することでオフロードする．この際に，オフロードできる処理が即高速化につながるかやコスト効果が十分かは分からないので，検証環境での性能測定を通じて，オフロードするしないを試行することで，より高速となるオフロードパターンを抽出する．

ここで，以前の研究で行っていたループ文のオフロードについては，個々のループ文検知を構文解析ツールで行い，そのループ文を GPU や FPGA にオフロードすることは OpenACC の #pragma を使った指定や OpenCL で定義された指示句を用いることでできた．しかし，機能ブロックのオフロードについては，機能ブロックの発見，その機能ブロックがオフロード用の既存ライブラリ/IP コア等があるかを発見，機能ブロックをライブラリ/IP コア等と置換した際にホスト側とのインタフェースを整合，の 3 つを考慮する必要がある．

### 3.4　機能ブロックのオフロードの処理方式

前サブ節の 3 つの考慮点に従い，機能ブロックオフロードの処理について詳細検討する．

A. 機能ブロックの発見

処理 A-1：構文解析にて，ソースコードから外部のライブラリの関数呼び出しを行っていることを検知する．FFT 等算術計算等のライブラリの呼び出しを検知することを想定しており，事前にコードパターン DB に外部ライブラリリストを保持しており，それとの照合で検知する．

処理 A-2：登録されていないライブラリ呼び出し以外の機能処理を検出するため，構文解析にてソースコードの定義記述からクラス，構造体等を検出する．

B. オフロード可能機能の発見

処理 B-1：特定のライブラリ，機能ブロックを高速化する GPU 用ライブラリや FPGA 用 IP コアとそれに関連する情報を，コードパターン DB に保持しておく．置換元のライブラリ，機能ブロックについては，機能名とともにコードや実行ファイルを登録する．A-1 で検出したライブラリ呼び出しに対して，ライブラリ名をキーに，高速化できる GPU 用ライブラリや FPGA 用 IP コアがあるかを検索する．

処理 B-2：処理 B-1 で，コードパターン DB に登録されている情報を利用する．処理 A-2 で検出した，クラス，構造体等の機能処理に対して，高速化するライブラリや IP コアがあるかを，類似性検出ツールで検出する．類似性検出ツールとは，Deckard 等，コピーコードやコピー後変更したコードの検出を対象とするツールであり，行列計算のコード等，CPU で計算する場合は記述が同様になる処理や，他者のコードをコピーして変更した処理等を一部検出できると考える．類似性検出ツールは，新規に独立に作成したようなクラス等については検出が困難となるため対象外である．コードパターン DB に登録された，特定の処理を高速化するライブラリや IP コアがある機能について，類似性が高いことをツールの閾値等で判定する．類似性検出でカバーできる範囲は 100%でないことは明らかなため，人工知能処理のパターン認識等の適用も将来的な応用として考えられる．教師有学習でパターン認識で頻繁に活用される SVM や，教師無学習で応用が広がっている Deep Learning の適用が想定される．ただし，人工知能処理の場合も，機械にコードの意図を理解させることは困難なため，新規独立に作成したようなクラス，構造体等については，対象外となると考える．

C. ホスト側プログラムとのインタフェース整合

処理 C-1：A-1 で検出したライブラリ呼び出しに対して，B-1 で該当するライブラリや IP コアを検索しているため，その置換するライブラリや IP コアを GPU や FPGA に実装し，ホスト側（CPU）プログラムと繋ぐ．ここで，GPU 用ライブラリの場合は，CUDA 等のライブラリを想定しており，C 言語コードから CUDA ライブラリを利用する手法がライブラリとともに公開されているため，コードパターン DB にもライブラリ利用手法も含めて登録しておき，その手法に従って利用する．FPGA 用 IP コアの場合は HDL 等が想定されるが，IP コア関連の情報として OpenCL コードもコードパターン DB に保持する．OpenCL コードから，OpenCL インタフェースを用いた CPU と FPGA の接続及び，FPGA への IP コア実装が，Xilinx や Intel 等の FPGA ベンダの高位合成ツール（Xilinx Vivado, Intel HLS Compiler 等）を介して行うことができる．

処理 C-2：A-2 で検出したクラス，構造体等に対して，B-2 で高速化できるライブラリや IP コアを検索しているため，その該当するライブラリや IP コアを GPU や FPGA に実装する．ここで，C-1 では，特定のライブラリ呼び出しに対して高速化するライブラリや IP コアであるため，インタフェース部分の生成等は必要になるが，GPU，FPGA とホスト側プログラムの想定する引数や戻り値の数や型は合っていた．しかし，B-2 は類似性等で判断しているため，引数や戻り値の数や型等の基

本的な部分があっている保証はない．

合っていない場合は，ライブラリや IP コアは既存ノウハウであり変更が頻繁にできるものでないため，オフロードを依頼するユーザに対して，元のコードの引数や戻り値の数や型について，ライブラリや IP コアに合わせて変更するか確認し，確認了承後にオフロード性能試験を試行する．型の違いについて，float と double 等キャストすればよいだけであれば，特にユーザ確認せずに試行に入ってもよい．また，引数や戻り値で，元のプログラムとライブラリや IP コアで数が異なる場合に，例えば，CPU プログラムで引数 1, 2 が必須で 3 がオプションであり，ライブラリや IP コアで引数 1, 2 が必須の場合等，省略しても問題ないような場合は，ユーザに確認せず，オプション引数は自動で無しとして扱うなどしてもよい．なお，引数や戻り値の数や型が完全に合っている場合は，C-1 と同様でよい．

## 4. 実　　装

### 4.1 利用ツール

3 節提案技術の有効性を確認するための実装を説明する．機能ブロックの GPU，FPGA 自動オフロードの有効性確認のため，対象アプリケーションは C/C++言語のアプリケーションとし，GPU は NVIDIA Quadro P4000(CUDA core: 1792, Memory: GDDR5 8GB)，FPGA は Intel PAC with Intel Arria10 GX FPGA を用いる．なお，FPGA にコンパイルするマシンは，DELL EMC PowerEdge R740（CPU : Intel Xeon Bronze 3104 / 1.70GHz，RAM : 32GB RDIMM DDR4-2666 × 2）である．

GPU 処理は市中の PGI コンパイラ 19.4 を用いる．PGI コンパイラは OpenACC を解釈する C/C++/Fortran 向けコンパイラであり，for 文等のループ文を，OpenACC のディレクティブ #pragma acc kernels, #pragma acc parallel loop で指定することにより，GPU 向けバイトコードを生成し，実行により GPU オフロードを可能としている．合わせて，cuFFT や cuBLAS 等の CUDA ライブラリの呼び出しも処理が可能である．

FPGA 処理は，Intel Acceleration Stack Version 1.2（Intel FPGA SDK for OpenCL 17.1.1，Quartus Prime Version 17.1.1）を用いる．Intel FPGA SDK for OpenCL は，標準 OpenCL に加え Intel 向けの#pragma 等を解釈する高位合成ツール（HLS）であり，FPGA で処理するカーネルと CPU で処理するホストプログラムを記述した OpenCL コードを，解釈しリソース量等の情報を出力し，FPGA の配線作業等を行い，FPGA で動作できるようにする．FPGA 実機で動作できるようにするには，100 行程度の小プログラムでも 3 時間程の長時間がかかるが，リソース量オーバーの際は早めにエラーとなる．FPGA の既存 OpenCL コードをカーネルコードに組み込めば，OpenCL プログラム処理の中でオフロードが可能である．

コードパターン DB は，MySQL8.0 を用いる．呼び出しているライブラリ名をキーに，高速化できる GPU 用ライブラリや FPGA 用 IP コアを検索するためのレコードを保持する．ライブラリや IP コアには，それに紐づく名前やコードや実行ファイルが保持される．実行ファイルはその利用手法等も登録されている．合わせて，ライブラリや IP コアを類似性検出技術で検出するための，比較用コードとの対応関係も保持される．

類似性検出ツールには，Deckard v2.0 [45] を用いる．Deckard は機能ブロックのオフロードの適用領域拡大のため，ライブラリ呼び出し以外にも，コードコピーし変更した機能等のオフロードを実現するため，照合対象となる部分コードと，DB に登録されたコードの類似性を判定する．

実装は C 言語で行い，次サブ節の処理を行う．

### 4.2 実 装 動 作

実装の動作概要を説明する．実装は，C/C++アプリケーションの利用依頼があると，まず，C/C++アプリケーションのコードを解析して，以前の研究であるループ文オフロードに使うためループ文やその回数を検出するとともに，A-1 呼び出されているライブラリや A-2 定義されているクラス，構造体等のプログラム構造を把握する．構文解析には，LLVM/Clang [46] の構文解析ライブラリ (libClang の python binding) を使う．呼び出されている外部ライブラリがあるかどうかは，コードパターン DB の外部ライブラリリストと照合することで確認する．

実装は，次に，B-1 呼び出されているライブラリを高速化できる GPU 用ライブラリ，FPGA 用 IP コアの検出を行う．呼び出されているライブラリをキーに，コードパターン DB に登録されているレコードから，高速化可能な実行ファイルや OpenCL 等を取得する．高速化できる置換用機能が見つかったら，実装は次に，C-1 その実行用ファイルを作成する．GPU 用ライブラリの場合は，置換用 CUDA ライブラリを呼び出すよう，C/C++コードに，元の部分は削除して置換記述する．FPGA 用 IP コアの場合は，取得した OpenCL コードを，元の部分をホストコードから削除してから，カーネルコードに置換記述する．それぞれ，置換記述が終わったら，GPU 向けには PGI コンパイラ，FPGA 向けには Intel Acceleration Stack でコンパイルする．FPGA に関しては，OpenCL コードに基づき，CPU と FPGA の接続が Intel の高位合成ツールを介して行われる．

ライブラリ呼び出しの場合について記載したが，類似性検知を用いる場合も並行して処理がされる．実装は，B-2 Deckard を用いて，検出されたクラス，構造体等の部分コードと DB に登録された比較用コードの類似性検知を行い，閾値越えの機能ブロックと該当する GPU 用ライブラリや FPGA 用 IP コアを発見する．B-1 の場合と同様に実行ファイルや OpenCL を取得する．実装は次に C-1 の場合と同様に実行用ファイルを作成するが，特に置換元のコードと置換するライブラリや IP コアの引数や戻り値，型等のインタフェースが異なる場合は，オフロードを依頼したユーザに対して，置換先ライブラリや IP コアに合わせて，インタフェースを変更してよいか確認し，確認後に実行用ファイルを作成する．

この時点で，検証環境の GPU や FPGA で性能測定できる実行用ファイルが作成される．機能ブロックオフロードについては，置換する機能ブロックが一つの場合は，その一つをオフロードするかしないかだけだが，複数ある場合は，一つづつオ

フロードするしないを検証パターンとして作成し，性能を測定し高速な解を発見する．これは，既存ノウハウとして高速化可能とされていても実測してみないとその条件で高速になるかわからないためである．例えば，5 つオフロード可能な機能ブロックがあり，実測の結果，2 番と 4 番のオフロードが高速化できた場合は，2 番と 4 番両方をオフロードするパターンで再度測定を行い，2 番と 4 番単独でオフロードする場合より高速となっている場合は，解として選択する．

このように，A, B，C の処理を中心に，全体としては，ソースコードを解析して，機能ブロックを発見し，置換可能な機能を発見し，ホスト側とのインタフェースを作成し，1 以上のオフロードパターンを検証環境で性能測定を行い，高速なパターンを見つける．

ここで，機能ブロックをオフロードすることによる性能向上がどの程度かは重要であり 5 節で検証する．また，A-1, B-1, C-1 の処理が基本となるが，より対象を増やすため，A-2, B-2, C-2 でどの程度，機能ブロックをオフロードできるかも 5 節で検証する．なお，実装について，FPGA の機能ブロックオフロードは未実装の部分があるため，5 節の評価では GPU だけ行う．

## 5. 評　　価

### 5.1 評価条件

#### 5.1.1 評価対象

評価対象は，IoT 等で多くのユーザが利用すると想定されるフーリエ変換と行列計算処理とする．

フーリエ変換処理は，振動周波数の分析等，IoT でのモニタリングの様々な場面で利用されている．IoT で，デバイスからデータをネットワーク転送するアプリケーションを考えた際に，ネットワークコストを下げるため，デバイス側で FFT 処理等の一次分析をして送ることは想定される．評価で用いる FFT は 2048 × 2048 のデータの 2 次元 FFT 処理を行い，FFT 処理高速化のため，CUDA の既存ライブラリ cuFFT [47] に自動置換する事で高速化する．

行列計算処理は，機械学習分析の様々な場面で利用されている．IoT，AI の普及により，クラウドだけでなくデバイス等様々な場面で行列計算が用いられることは多いため，既存アプリケーション含めて自動高速化のニーズはある．行列計算処理では，2048 × 2048 の直行行列データの LU 分解処理を行い，高速化のため，CUDA の既存ライブラリ cuSOLVER [48] に自動置換することで高速化する．

フーリエ変換，行列計算処理とも，CPU 向けのオリジナルコードは，Numerical Recipes in C [49] に記載のコードを用いる．

#### 5.1.2 評価手法

機能ブロックの GPU，FPGA オフロードについては，実利用の際はループ文オフロードと組み合わせて行う．しかし，ループ文については以前研究 [33] 等で評価しているため，本節では機能ブロックの GPU オフロードだけ評価する．対象のアプリケーションに対して，オフロード可能な機能ブロックを事前に

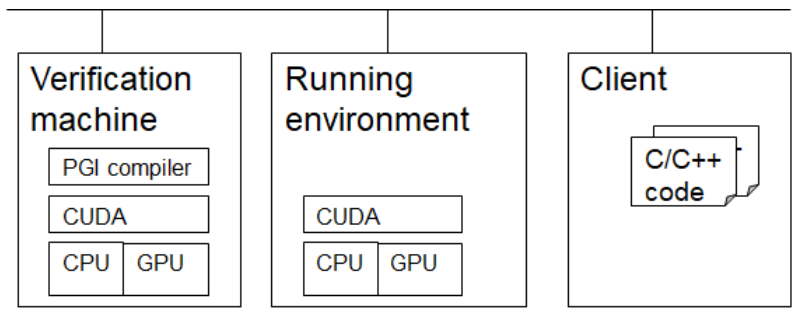

| Name | Hardware | CPU | RAM | GPU | OS | CUDA toolkit | PGI Compiler | MySQL |
|---|---|---|---|---|---|---|---|---|
| Verification machine | Dell Precision Tower 3620 | Intel(R) Core(TM) i5-7500 CPU@3.40GHz | 32GB | NVIDIA Quadro P4000 (CUDA core: 1792, Memory: GDDR5 8GB) | Ubuntu 16.04.6 LTS | 10.1 | 19.4 | 8.0 |
| Running environment | Dell Precision Tower 3620 | Intel(R) Core(TM) i5-7500 CPU@3.40GHz | 32GB | NVIDIA Quadro P4000 (CUDA core: 1792, Memory: GDDR5 8GB) | Ubuntu 16.04.6 LTS | 10.1 | | |
| Client | HP ProBook 470 G3 | Intel Core i5-6200U @2.3GHz | 8GB | | Windows 7 Professional | | | |

図 3　性能測定環境

DB に準備しておき，それに自動置換した際の性能を測定する．

実験の条件は以下で行う．

オフロード元：フーリエ変換アプリケーション，行列計算処理アプリケーション

オフロード先：cuFFT，cuSOLVER

オフロード元発見手法：オフロード元アプリケーションのコードは，コード側で外部ライブラリを呼んでおり DB の名前照合で発見する場合と，コード側でライブラリ処理をコピーしてコメントを入れており類似性検出ツールで発見する場合の 2 パターン準備．

比較対象：提案手法，全て CPU 処理，ループ文自動オフロード手法．ループ文自動オフロード手法は，[33] で，ループ文を検知して GPU オフロードに適したループ文のパターンを，遺伝的アルゴリズムを用いて検証環境で繰返し測定することで，段々と高速のパターンを探索する手法である．

性能測定：フーリエ変換ではグリッドサイズ 2048 × 2048 でサンプルテスト処理，行列計算処理では 2048 × 2048 の直行行列の LU 分解処理の処理時間を測定し，比較対象と比較．

#### 5.1.3 評価環境

利用する GPU として NVIDIA Quadro P4000 を備えた物理マシンを検証に用いる．NVIDIA Quadro P4000 の CUDA コア数は 1792 である．PGI コンパイラはコミュニティ版の 19.4，CUDA Toolkit は 10.1 を用いる．評価環境とスペックを図 3 に示す．ここで，Client ノート PC から，ユーザが利用する C/C++言語アプリケーションコードを指定し，Verification Machine を用いてチューニング後，実際のユーザが使う Running environment にデプロイする形である．

### 5.2 性能結果

IoT 等で多くのユーザが利用すると想定されるアプリケーションとして，フーリエ変換と行列計算処理の高速化を確認した．

図 4 は以前研究 [33] でフーリエ変換の GPU での自動高速化を行った際の例である．各世代個体の最高性能と GA の世代数をグラフにとり，性能は CPU のみで処理の場合との比で示している．全て CPU 処理と比べて GPU にオフロードして 5 倍以上の性能が実現出来ていることが分かる．

以前の結果を踏まえ，今回提案手法の実装により，どの程度性能が改善されたかの測定結果を示す．まず，オフロード元発見手法は，ライブラリ呼び出ししている場合でも，コードをコ

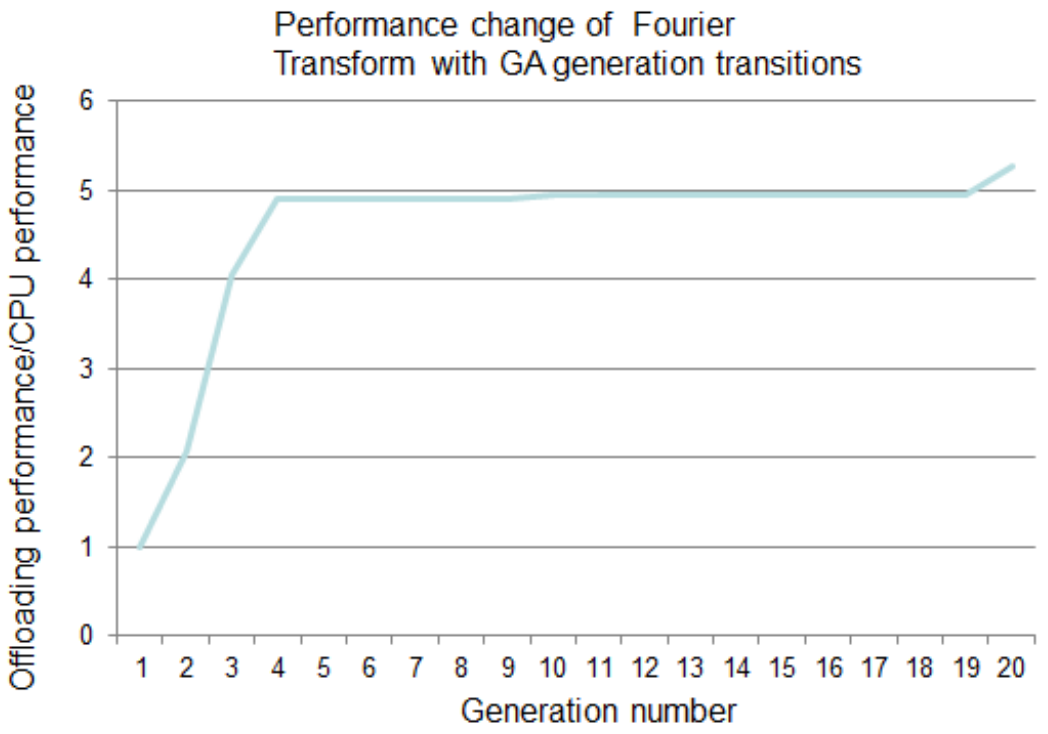

図 4 参考：以前研究での GA 世代数に伴うフーリエ変換の性能変化 [33]

| | Performance improvement of loop offloading [33] | Performance improvement of function blocks offloading |
|---|---|---|
| Fourier transform | 5.4 | 730 |
| Matrix calculation | 38 | 130000 |

図 5 以前研究と今回提案手法での性能改善度の比較

ピーしている場合でも，提案の DB 照合及び類似性検出ツールにより，置換できている．図 5 は，機能ブロックオフロードした際の性能が，全 CPU 処理に比べて何倍になっているかを示した表であり，比較対象である [33] のループ文自動オフロードでの性能改善も合わせて示している．図 5 より，フーリエ変換については，以前のループ文オフロードでは 5.4 倍に留まっていたのが，今回提案手法により 730 倍の性能が実現できていることがわかる．行列計算処理については，以前研究では 38 倍に留まっていたのが，今回提案手法により 130,000 倍の性能が実現できていることがわかる．また，以前研究では，ループ文探索に GA を用いるため数時間以上探索にかかったが，今回の機能ブロックのオフロードについては数分で処理が完了している．

## 6. ま と め

本稿では，私が提案している，ソフトウェアを配置先環境に合わせて自動適応させ FPGA，GPU 等を適切に利用して，アプリケーションを高性能に運用するための環境適応ソフトウェアの要素として，機能ブロックの GPU，FPGA へのオフロード自動化手法を提案し，評価した．

提案した機能ブロックオフロード手法は，ソースコードの分析から始まる．ソースコードを分析し，オフロード可能ライブラリ呼び出しを DB 照合で検知し，DB に登録された置換可能な GPU 用ライブラリや FPGA 用 IP コアの利用に置換する．置換された GPU，FPGA 用機能を含めた形で，検証環境で性能測定を行い，最高性能のパターンを解とする．ソースコード分析では，より多くの置換可能機能ブロックを検索するため，類似性検知技術を用いた，オフロード可能機能ブロックの検索も行い，同様手法での置換，性能測定を行う．ただし，置換可能と判定しても，引数や戻り値のインタフェースが異なる場合は，置換可能機能の方のインタフェースと合わせて良いかを利用者に確認する．

既存アプリケーションに対して提案手法での GPU 自動オフロードを行い，方式の有効性を確認した．今後は，FPGA へのオフロードやより多くのアプリケーションでの評価を行う．オフロード可能な機能ブロック，ループ文を含む一般的アプリケーションを，FPGA，GPU 両方使える環境で適切にオフロードする手法も検討する．

## 文 献

[1] A. Putnam, et al., "A reconfigurable fabric for accelerating large-scale datacenter services," ISCA'14, pp.13-24, 2014.
[2] O. Sefraoui, et al., "OpenStack: toward an open-source solution for cloud computing," International Journal of Computer Applications, Vol.55, 2012.
[3] Y. Yamato, et al., "Fast and Reliable Restoration Method of Virtual Resources on OpenStack," IEEE Transactions on Cloud Computing, Sep. 2015.
[4] Y. Yamato, et al., "Software Maintenance Evaluation of Agile Software Development Method Based on OpenStack," IEICE Transactions on Information & Systems, Vol.E98-D, No.7, pp.1377-1380, July 2015.
[5] Y. Yamato, "Automatic verification technology of software patches for user virtual environments on IaaS cloud," Journal of Cloud Computing, Springer, 2015, 4:4, Feb. 2015.
[6] Y. Yamato, "Optimum Application Deployment Technology for Heterogeneous IaaS Cloud," Journal of Information Processing, Vol.25, No.1, pp.56-58, Jan. 2017.
[7] Y. Yamato, "Performance-Aware Server Architecture Recommendation and Automatic Performance Verification Technology on IaaS Cloud," Service Oriented Computing and Applications, Springer, Nov. 2016.
[8] Y. Yamato, "Server Selection, Configuration and Reconfiguration Technology for IaaS Cloud with Multiple Server Types," Journal of Network and Systems Management, Springer, Aug. 2017.
[9] Y. Yamato, et al., "Evaluation of Agile Software Development Method for Carrier Cloud Service Platform Development," IEICE Transactions on Information & Systems, Vol.E97-D, No.11, pp.2959-2962, Nov. 2014.
[10] Y. Yamato, "Server Structure Proposal and Automatic Verification Technology on IaaS Cloud of Plural Type Servers," International Conference on Internet Studies (NETs2015), July 2015.
[11] Y. Yamato, "Proposal of Optimum Application Deployment Technology for Heterogeneous IaaS Cloud," 2016 6th International Workshop on Computer Science and Engineering (WCSE 2016), pp.34-37, June 2016.
[12] Y. Yamato, "Cloud Storage Application Area of HDD-SSD Hybrid Storage, Distributed Storage and HDD Storage," IEEJ Transactions on Electrical and Electronic Engineering, Vol.11, Issue.5, pp.674-675, Sep. 2016.
[13] Y. Yamato, "Use case study of HDD-SSD hybrid storage, distributed storage and HDD storage on OpenStack," 19th International Database Engineering & Applications Symposium (IDEAS15), pp.228-229, July 2015.
[14] Y. Yamato, "OpenStack Hypervisor, Container and Baremetal Servers Performance Comparison," IEICE Communication Express, Vol.4, No.7, pp.228-232, July 2015.
[15] AWS EC2 web site, https://aws.amazon.com/ec2/instance-types/
[16] J. E. Stone, et al., "OpenCL: A parallel programming standard for heterogeneous computing systems," Computing in science & engineering, Vol.12, No.3, pp.66-73, 2010.
[17] J. Sanders and E. Kandrot, "CUDA by example : an introduction to general-purpose GPU programming," Addison-Wesley, 2011

[18] M. Hermann, et al., "Design Principles for Industrie 4.0 Scenarios," Rechnische Universitat Dortmund. 2015.
[19] Y. Yamato, et al., "Predictive Maintenance Platform with Sound Stream Analysis in Edges," Journal of Information Processing, Vol.25, pp.317-320, Apr. 2017.
[20] Y. Yamato, Y. Fukumoto and H. Kumazaki, "Proposal of Real Time Predictive Maintenance Platform with 3D Printer for Business Vehicles," 5th International Conference on Software and Information Engineering (ICSIE 2016), May 2016.
[21] Tron project web site, http://www.tron.org/
[22] P. C. Evans and M. Annunziata, "Industrial Internet: Pushing the Boundaries of Minds and Machines," Technical report of General Electric (GE), Nov. 2012.
[23] Y. Yamato, "Ubiquitous Service Composition Technology for Ubiquitous Network Environments," IPSJ Journal, Vol.48, No.2, pp.562-577, Feb. 2007.
[24] M. Takemoto, et al., "Service-composition Method and Its Implementation in Service-provision Architecture for Ubiquitous Computing Environments," IPSJ Journal, Vol.46, No.2, pp.418-433, Feb. 2005. (in Japanese)
[25] M. Takemoto, et al., "Service Elements and Service Templates for Adaptive Service Composition in a Ubiquitous Computing Environment," The 9th Asia-Pacific Conference on Communications (APCC2003), Vol.1, pp.335-338, Sep. 2003.
[26] Y. Yamato, et al., "Study of Service Processing Agent for Context-Aware Service Coordination," IEEE International Conference on Service Computing (SCC 2008), pp.275-282, July 2008.
[27] Y. Yamato, et al., "Development of Service Control Server for Web-Telecom Coordination Service," IEEE International Conference on Web Services (ICWS 2008), pp.600-607, Sep. 2008.
[28] H. Sunaga, et al., "Service Delivery Platform Architecture for the Next-Generation Network," ICIN2008, Oct. 2008.
[29] Y. Yamato, et al., "Study and Evaluation of Context-Aware Service Composition and Change-Over Using BPEL Engine and Semantic Web Techniques," IEEE Consumer Communications and Networking Conference (CCNC 2008), pp.863-867, Jan. 2008.
[30] H. Sunaga, et al., "Ubiquitous Life Creation through Service Composition Technologies," World Telecommunications Congress 2006 (WTC2006), May 2006.
[31] J. Gosling, et al., "The Java language specification, third edition," Addison-Wesley, 2005. ISBN 0-321-24678-0.
[32] Y. Yamato, et al., "Automatic GPU Offloading Technology for Open IoT Environment," IEEE Internet of Things Journal, Sep. 2018.
[33] Y. Yamato, "Study of parallel processing area extraction and data transfer number reduction for automatic GPU offloading of IoT applications," Journal of Intelligent Information Systems, Springer, DOI:10.1007/s10844-019-00575-8, 2019.
[34] K. Shirahata, et al., "Hybrid Map Task Scheduling for GPU-Based Heterogeneous Clusters,"IEEE CloudCom, 2010.
[35] Altera SDK web site, https://www.altera.com/products/design-software/embedded-software-developers/opencl/documentation.html
[36] Xilinx SDK web site, https://japan.xilinx.com/html_docs/xilinx2017_4/sdaccel_doc/lyx1504034296578.html
[37] S. Wienke, et al., "OpenACC-first experiences with real-world applications," Euro-Par Parallel Processing, 2012.
[38] M. Wolfe, "Implementing the PGI accelerator model," ACM the 3rd Workshop on General-Purpose Computation on Graphics Processing Units, pp.43-50, Mar. 2010.
[39] K. Ishizaki, "Transparent GPU exploitation for Java," CANDAR 2016, Nov. 2016.
[40] E. Su, et al., "Compiler support of the workqueuing execution model for Intel SMP architectures," In Fourth European Workshop on OpenMP, Sep. 2002.
[41] Jenkins web site, https://jenkins.io/
[42] Selenium web site, https://www.seleniumhq.org/
[43] J. H. Holland, "Genetic algorithms," Scientific american, Vol.267, No.1, pp.66-73, 1992.
[44] OpenCV web site, http://opencv.org/
[45] Deckard web site, http://github.com/skyhover/Deckard
[46] Clang website, http://llvm.org/
[47] cuFFT website, https://docs.nvidia.com/cuda/cufft/index.html
[48] cuSOLVER website, https://docs.nvidia.com/cuda/cusolver/index.html
[49] Numerical Recipes in C, https://www.cec.uchile.cl/cinetica/pcordero/MC_libros/NumericalRecipesinC.pdf